\documentclass[12pt]{article}
\usepackage[textures]{graphicx}  
\begin{document}

\small{.}\vspace{15mm}

\begin{center}{\bf \Huge GRAVITY} 

\vspace{5mm}

{\bf \Large AS}

\vspace{5mm}

{\bf \Huge  QUANTUM FOAM IN-FLOW } 
\vspace{33mm}

{{\bf \LARGE Reginald T. Cahill}}

\vspace{20mm}

      {\bf \Large School of Chemistry, Physics and Earth Sciences}
\vspace{7mm}

      {\bf \Large  Flinders University }
\vspace{7mm}

      {\bf \large GPO Box 2100, Adelaide 5001, Australia }
\vspace{15mm}

      {\bf \large Reg.Cahill@flinders.edu.au}
\vspace{5mm}

{\bf Process Physics URL: }
      
{\bf http://www.scieng.flinders.edu.au/cpes/people/cahill\_r/processphysics.html}

\vspace{5mm}

{\bf - June 2003 -}

\end{center}
\newpage

\vspace{10mm}
\begin{center}
{\hspace*{800mm} \\ \bf \large  Abstract}
\end{center}
\vspace{10mm}
{\small The new information-theoretic {\it Process Physics}  provides an
explanation of space as a quantum foam  system in which gravity is an
inhomogeneous flow of the quantum foam into matter.   The older Newtonian and
General Relativity theories for gravity are analysed.  It is shown that 
Newtonian gravity may be written in the form of an in-flow.  General Relativity
is also analysed as an in-flow, for those cases where it has been tested. An
analysis of various  experimental data demonstrates that absolute motion  relative
to space  has been observed  by Michelson and Morley, Miller, Illingworth,  Jaseja
{\it et al}, Torr and Kolen, and by  DeWitte.  The  Dayton Miller  and Roland
DeWitte data also reveal the in-flow of space into matter which manifests as 
gravity. The experimental data suggests that the in-flow is turbulent, which
amounts to the observation of a gravitational  wave phenomena.  A new in-flow
theory of gravity is proposed which passes all the tests that General Relativity 
was claimed to have passed, but as well the new theory suggests that the
so-called spiral galaxy rotation-velocity anomaly may be explained without the
need of `dark matter'.  Various other gravitational anomalies also appear to be
explainable. Newtonian gravity appears to be strictly valid only outside of  spherically
symmetric matter systems. }

\vspace{25mm}
  Key words:  Process Physics,   quantum foam,  quantum

  gravity, absolute motion, dark matter, stellar structure, 
 
  gravitational anomalies, general relativity, Newtonian gravity

\newpage

 \tableofcontents
\newpage

\section{  Introduction\label{section:introduction}}

The new information-theoretic {\it Process Physics}
\cite{RCPP2003,RCAMGE,RC01,RC02,CK97,CK98,CK99,CKK00,CK,RC03,RC05,Kitto,MC}  provides for the first
time an explanation of space as a quantum foam system in which gravity is an inhomogeneous flow of the
quantum foam into matter.  An analysis \cite{RCPP2003,RCAMGE} of data from various experiments
demonstrates that absolute motion  relative to space  has been observed  by Michelson and Morley
\cite{MM}, Miller
\cite{Miller2}, Illingworth \cite{Illingworth}, Jaseja {\it et al}
\cite{Jaseja}, Torr and Kolen
\cite{Torr}, and by  DeWitte \cite{DeWitte}, contrary to common belief  within physics that absolute
motion has never been observed.   The  Dayton Miller and Roland DeWitte data also reveal the in-flow of
space into matter which manifests as  gravity.    The experimental data suggests that the in-flow manifest
turbulence, which amounts to the observation of a gravitational wave phenomena.  The Einstein assumptions
leading to the Special and General Theory of Relativity are shown to be falsified by the extensive
experimental data.

Contrary to the Einstein assumptions absolute motion is consistent with relativistic effects, which are
caused by actual dynamical effects of absolute motion through the quantum foam.  Lorentzian 
relativity  is seen to be essentially  correct.  Vacuum Michelson interferometer experiments or its
equivalent \cite{vacuum, KT, BH, Muller, NewVacuum} cannot detect absolute motion, though their null
results are always incorrectly reported as evidence of an absence of absolute motion.

A new  in-flow theory of gravity in the classical limit is proposed. It passes all the standard tests of both
the Newtonian and the General Relativity  theories of gravity.  However it appears that this new theory may
explain the spiral galaxy rotation-velocity anomaly without invoking dark matter.   As well this new theory is
expected to predict turbulent flow, and such behaviour is manifest in the existing experimental
observations of absolute motion.  Other gravitational anomalies also now appear to be capable of being
explained.

This paper is a condensed version of certain sections of Cahill \cite{RCPP2003}.

\section{ A New Theory of Gravity  \label{section:gravity}}

\subsection{ Classical Effects of Quantum Foam \label{subsection:classicalquantum}}

We begin here the analysis that will lead to the new theory and explanation of gravity.  In this theory
gravitational effects are caused solely by an inhomogeneous flow of the quantum foam. The new
information-theoretic concepts underlying this physics were discessed in \cite{RCPP2003, RC01, RC02}. 
Essentially matter effectively acts as a `sink' for that quantum foam.  To begin with it should be noted
that even Newtonian gravity is suggestive of a flow explanation of gravity. In that theory the
gravitational acceleration
${\bf g}$ is determined by  the matter density
$\rho$ according to
\begin{equation}\label{eqn:g1}
\nabla.{\bf g}=-4\pi G\rho.
\end{equation}
For $\nabla \times {\bf g}=0$ this gravitational acceleration ${\bf g}$ may be written as the
gradient of the gravitational potential $\Phi$ 
\begin{equation}{\bf g}=-{\bf \nabla}\Phi,\end{equation}  
where the  gravitational
potential is now determined  by $ \nabla^2\Phi=4\pi G\rho $.  Here, as usual, $G$ is the gravitational constant. 
Now as $\rho\geq 0$ we can choose to have 
$\Phi
\leq 0$ everywhere if $\Phi \rightarrow 0$ at infinity. So we can introduce  ${\bf v}^2=-2\Phi \geq 0$ where 
$\bf v$ is some velocity vector field.
  Here the value of ${\bf v}^2$ is
specified, but not the direction of ${\bf v}$. Then
\begin{equation}
{\bf g}=\frac{1}{2}{\bf \nabla}({\bf v}^2)=({\bf v}.{\bf \nabla}){\bf v}+
{\bf v}\times({\bf \nabla}\times{\bf v}).
\label{eqn:f1}
\end{equation} 
For irrotational flow 
${\bf \nabla} \times {\bf v}={\bf 0}$. Then ${\bf g}$ is
the usual Euler expression for the  acceleration of a fluid element in a
time-independent or stationary fluid flow.   If the flow is time dependent that expression
is expected to become
 \begin{equation}{\bf g}=({\bf v}.{\bf \nabla}){\bf v}+{\bf v}\times({\bf \nabla}\times{\bf
v})
+\displaystyle{\frac{\partial {\bf v}}{\partial t}}.
\label{eqn:f2}\end{equation}
This equation  is then to be accompanied by the `Newtonian equation' for the flow field
\begin{equation}
\frac{1}{2}\nabla^2({\bf v}^2)=-4\pi G\rho.
\label{eqn:f3}\end{equation}
While this hints at a fluid flow interpretation of  Newtonian gravity  the fact that the direction of ${\bf v}$ is
not specified by (\ref{eqn:f3}) suggests that  some generalisation  is to be expected in which the
direction of ${\bf v}$ is specified.  Of course within the fluid flow interpretation (\ref{eqn:f2}) and
(\ref{eqn:f3}) are together equivalent to the Universal Inverse Square Law for Gravity.
Indeed for a spherically symmetric distribution of matter of total mass $M$ the velocity field outside of the
matter 
\begin{equation}
{\bf v}({\bf r})=-\sqrt{\frac{2GM}{r}}\hat{\bf r},
\label{eqn:vfield}\end{equation}
satisfies (\ref{eqn:f3}) and reproduces the inverse square law form for ${\bf g}$ using (\ref{eqn:f2}): 
\begin{equation}
{\bf g}=-\frac{GM}{r^2}\hat{\bf r}.
\label{eqn:InverseSqLaw}\end{equation} 
The in-flow direction
$-\hat{\bf r}$  in (\ref{eqn:vfield}) may be replaced by any other direction, in which case however
the direction of ${\bf g}$ in (\ref{eqn:InverseSqLaw})  remains radial. 

Of the many new effects
predicted by the generalisation of (\ref{eqn:f3}) one is that this `Inverse Square Law'  is only
valid outside of spherically symmetric matter systems.  Then, for example, the `Inverse Square Law'
is expected to be inapplicable to spiral galaxies. The incorrect assumption of the universal validity
of this law led to the notion of  `dark matter' in order to reconcile the faster observed rotation 
velocities of matter within  such galaxies compared to that predicted by the above law.

To arrive at the new in-flow theory of gravity we require that the velocity field  ${\bf v}({\bf r},t)$ be
specified and measurable with respect to a suitable frame of reference.  We shall use the Cosmic Microwave
Background (CMB) frame of reference for that purpose \cite{CMB}.  Then an `object' has velocity  
 ${\bf v}_0(t)=d{\bf r}_0(t)/dt$ 
with respect to that CMB frame, where ${\bf r}_0(t)$  is the position of the object
wrt that frame.   We then define 
\begin{equation}
{\bf v}_R(t) ={\bf v}_0(t) - {\bf v}({\bf r}_0(t),t),
\label{eqn:18}\end{equation}
as the velocity of the object relative to the quantum foam at the location of the object.

Process Physics appears to be leading to the Lorentzian interpretation of so called `relativistic effects'.  This
means that the speed of light is only `c' wrt the quantum-foam system, and that time dilation effects for clocks
and length contraction effects for rods are caused by the motion of clocks and rods relative to the quantum foam.
So these effects are real dynamical effects caused by the quantum foam, and are not to be interpreted as
spacetime effects as suggested by Einstein.  To arrive at the dynamical description of the various effects of the
quantum foam we shall introduce conjectures that essentially lead to a phenomenological description of these
effects. In the future we expect to be able to derive this dynamics directly from the QHFT formalism. First we
shall conjecture that the path of an object through an inhomogeneous and time-varying quantum-foam is determined by
a variational principle, namely the path ${\bf r}_0(t)$ minimises the travel time  (for early investigations of
the in-flow approach to gravity see Ives\cite{Ives} and Kirkwood\cite{RK1,RK2}), 
\begin{equation}
\tau[{\bf r}_0]=\int dt \left(1-\frac{{\bf v}_R^2}{c^2}\right)^{1/2},
\label{eqn:f4}
\end{equation}  
with ${\bf v}_R$ given by (\ref{eqn:18}). Under a deformation of
the trajectory  ${\bf r}_0(t) \rightarrow  {\bf r}_0(t) +\delta{\bf r}_0(t)$,
${\bf v}_0(t) \rightarrow  {\bf v}_0(t) +\displaystyle\frac{d\delta{\bf r}_0(t)}{dt}$,  and we also
have
\begin{equation}\label{eqn:G2}
{\bf v}({\bf r}_0(t)+\delta{\bf r}_0(t),t) ={\bf v}({\bf r}_0(t),t)+(\delta{\bf
r}_0(t).{\bf \nabla}) {\bf v}({\bf r}_0(t))+... 
\end{equation}
Then
\begin{eqnarray}\label{eqn:G3}
\delta\tau&=&\tau[{\bf r}_0+\delta{\bf r}_0]-\tau[{\bf r}_0]  \nonumber\\
&=&-\int dt \:\frac{1}{c^2}{\bf v}_R. \delta{\bf v}_R\left(1-\displaystyle{\frac{{\bf
v}_R^2}{c^2}}\right)^{-1/2}+...\nonumber\\
&=&\int dt\frac{1}{c^2}\left({\bf
v}_R.(\delta{\bf r}_0.{\bf \nabla}){\bf v}-{\bf v}_R.\frac{d(\delta{\bf
r}_0)}{dt}\right)\left(1-\displaystyle{\frac{{\bf v}_R^2}{c^2}}\right)^{-1/2}+...\nonumber\\ 
&=&\int dt \frac{1}{c^2}\left(\frac{{\bf v}_R.(\delta{\bf r}_0.{\bf \nabla}){\bf v}}{ 
\sqrt{1-\displaystyle{\frac{{\bf
v}_R^2}{c^2}}}}  +\delta{\bf r}_0.\frac{d}{dt} 
\frac{{\bf v}_R}{\sqrt{1-\displaystyle{\frac{{\bf
v}_R^2}{c^2}}}}\right)+...\nonumber\\
&=&\int dt\: \frac{1}{c^2}\delta{\bf r}_0\:.\left(\frac{({\bf v}_R.{\bf \nabla}){\bf v}+{\bf v}_R\times({\bf
\nabla}\times{\bf v})}{ 
\sqrt{1-\displaystyle{\frac{{\bf
v}_R^2}{c^2}}}}  +\frac{d}{dt} 
\frac{{\bf v}_R}{\sqrt{1-\displaystyle{\frac{{\bf
v}_R^2}{c^2}}}}\right)+...
\end{eqnarray}
  Hence a 
trajectory ${\bf r}_0(t)$ determined by $\delta \tau=0$ to $O(\delta{\bf r}_0(t)^2)$ satisfies 
\begin{equation}\label{eqn:G4}
\frac{d}{dt} 
\frac{{\bf v}_R}{\sqrt{1-\displaystyle{\frac{{\bf v}_R^2}{c^2}}}}=-\frac{({\bf
v}_R.{\bf \nabla}){\bf v}+{\bf v}_R\times({\bf
\nabla}\times{\bf v})}{ 
\sqrt{1-\displaystyle{\frac{{\bf v}_R^2}{c^2}}}}.
\end{equation}
Let us now write this in a more explicit form.  This will
also allow the low speed limit to be identified.   Substituting ${\bf
v}_R(t)={\bf v}_0(t)-{\bf v}({\bf r}_0(t),t)$ and using 
\begin{equation}\label{eqn:G5}
\frac{d{\bf v}({\bf r}_0(t),t)}{dt}=({\bf v}_0.{\bf \nabla}){\bf
v}+\frac{\partial {\bf v}}{\partial t},
\end{equation}
we obtain
\begin{equation}\label{eqn:G6}
\frac{d}{dt} 
\frac{{\bf v}_0}{\sqrt{1-\displaystyle{\frac{{\bf v}_R^2}{c^2}}}}={\bf v}
\frac{d}{dt}\frac{1}{\sqrt{1-\displaystyle{\frac{{\bf v}_R^2}{c^2}}}}+\frac{({\bf v}.{\bf
\nabla}){\bf v}-{\bf v}_R\times({\bf
\nabla}\times{\bf v})+\displaystyle{\frac{\partial {\bf v}}{\partial t}}}{ 
\displaystyle{\sqrt{1-\frac{{\bf v}_R^2}{c^2}}}}.
\end{equation}
Then in the low speed limit  $v_R \ll c $   we  obtain
\begin{equation}{\label{eqn:G7}}
\frac{d{\bf v}_0}{dt}=({\bf v}.{\bf
\nabla}){\bf v}-{\bf v}_R\times({\bf \nabla}\times{\bf v})+\frac{\partial {\bf v}}{\partial t}={\bf g}({\bf
r}_0(t),t)+({\bf \nabla}\times{\bf v})\times{\bf v}_0,
\end{equation}
which agrees with the  `Newtonian' form (\ref{eqn:f2}) for zero vorticity (${\bf \nabla}\times{\bf
v}=0$).   Hence (\ref{eqn:G6}) is a generalisation of (\ref{eqn:f2}) to include  Lorentzian dynamical
effects, for 
in (\ref{eqn:G6})  we can multiply both sides by the rest mass  $m_0$ of the object, and   then
(\ref{eqn:G6}) involves 
\begin{equation}
m({\bf v}_R) =\frac{m_0}{\sqrt{1-\displaystyle{\frac{{\bf v}_R^2}{c^2}}}},
\label{eqn:G8}\end{equation}
the so called `relativistic' mass, and (\ref{eqn:G6}) acquires the form $\frac{d}{dt}(m({\bf v}_R){\bf
v}_0)={\bf F}$, where
${\bf F}$ is an effective `force' caused by the inhomogeneities and time-variation of the flow.  This is
essentially Newton's 2nd Law of Motion in the case of gravity only. That $m_0$ cancels is the equivalence principle, 
and which acquires a simple explanation in terms of the flow.  Note that the occurrence of
$1/\sqrt{1-\frac{{\bf v}_R^2}{c^2}}$ will lead to the precession of the perihelion of planetary orbits, and
also to horizon effects wherever  $|{\bf v}| = c$: the region where  $|{\bf v}| < c$ is
inaccessible from the region where $|{\bf v}|>c$.  Also (\ref{eqn:f4}) is easily used to determine the 
clock rate offsets in the GPS satellites, when the in-flow is given by (\ref{eqn:vfield}).

Eqn.(\ref{eqn:f4})  involves various absolute quantities such  as the absolute velocity of an object
relative to the  quantum foam and the absolute speed
$c$ also relative to the foam, and of course absolute velocities are excluded from the General Relativity (GR)
formalism.  However (\ref{eqn:f4}) gives (with $t=x_0^0$)
\begin{equation}
d\tau^2=dt^2-\frac{1}{c^2}(d{\bf r}_0(t)-{\bf v}({\bf r}_0(t),t)dt)^2=
g_{\mu\nu}(x_0 )dx^\mu_0 dx^\nu_0,
\label{eqn:24}\end{equation}
which is  the   Panlev\'{e}-Gullstrand form of the metric $g_{\mu\nu}$  \cite{PP, AG} for GR. All of the above
is very suggestive that  useful information for the flow dynamics may be obtained from GR by restricting the
choice of metric to the  Panlev\'{e}-Gullstrand form.  We emphasize that the absolute velocity ${\bf v}_R$ 
has been measured, and so the foundations of GR as usually stated are invalid. As we shall now see the GR
formalism involves two phenomena, namely (i) the use of an unnecessarily  restrictive Einstein measurement protocol and
(ii) the Lorentzian quantum-foam  dynamical effects.  Later we shall remove this measurement protocol from GR
and discover that the GR formalism reduces to explicit fluid flow  equations.   One significant implication of
this is that the whole spacetime formalism  introduced by Einstein evaporates - it is nothing more than an
{\it epicycle effect}, that is,  like Ptolemy's epicycles, it is an artifact and  arises from not being aware
of the influence of  certain features of a measurement procedure.

 However to understand  the GR formalism we
need to explicitly introduce the troublesome Einstein  measurement protocol and the peculiar effects that it
induces in the observers historical records, namely that they have a curved spacetime form. 
Again we emphasize that experimentally this measurement protocol is unnecessarily restrictive - one can do
measurements of absolute motion, and then the curvature disappears.

\subsection{  The Einstein Measurement Protocol\label{subsection:theeinsteinmeasurement}}

The quantum foam, it is argued, induces actual dynamical time dilations and length contractions in agreement
with the Lorentz interpretation of special relativistic effects.  Then observers in  uniform motion
`through' the foam will on measurement  of the speed of light obtain always the same numerical value  $c$.   To see
this explicitly consider how various observers $P, P^\prime,..$ moving with  different  speeds through the
foam, measure the speed of light.  They  each acquire a standard rod  and an accompanying standardised clock.
That means that these standard  rods  would agree if they were brought together, and at rest with respect to the
quantum foam they would all have length $\Delta l_0$, and similarly for the clocks.    Observer $P$ and
accompanying rod are both moving at  speed $v_R$ relative to the quantum foam, with the rod longitudinal to
that motion. P  then  measures the time
$\Delta t_R$, with the clock at end $A$ of the rod,  for a light pulse to travel from  end $A$ to the other end
$B$  and back again to $A$. The  light  travels at speed $c$ relative to the quantum-foam. Let the time taken for
the light pulse to travel from
$A\rightarrow B$ be $t_{AB}$ and  from $B\rightarrow A$ be $t_{BA}$, as measured by a clock at rest with respect
to the quantum foam\footnote{Not all clocks will behave in this same `ideal' manner.}. The  length of the rod 
moving at speed
$v_R$ is contracted to 
\begin{equation}
\Delta l_R=\Delta l_0\sqrt{1-\frac{v_R^2}{c^2}}.
\label{eqn:c0}\end{equation}
In moving from  $A$ to $B$ the light must travel an extra  distance 
because the  end  $B$ travels a distance $v_Rt_{AB}$ in this time, thus the total distance that must be
traversed  is
\begin{equation}\label{eqn:c1}
ct_{AB}=\Delta l_R+v_Rt_{AB},
\end{equation}
Similarly on returning from $B$ to $A$ the light must travel the distance
\begin{equation}\label{eqn:c2}
ct_{BA}=\Delta l_R-v_Rt_{BA}.
\end{equation}
Hence the total travel time $\Delta t_0$ is
\begin{eqnarray}\label{eqn:c3}
\Delta t_0=t_{AB}+t_{BA}&=&\frac{\Delta l_R}{c-v_R}+\frac{\Delta l_R}{c+v_R}\\
&=&\frac{2\Delta l_0}{c\sqrt{1-\displaystyle\frac{v_R^2}{c^2}}}.
\end{eqnarray}
Because  of  the time dilation effect for the moving clock
\begin{equation}
\Delta t_R=\Delta t_0\sqrt{1-\displaystyle\frac{v_R^2}{c^2}}.
\label{eqn:c4}\end{equation}
Then for the moving observer the speed of light is defined as the distance the observer believes the light
travelled ($2\Delta l_0$) divided by the travel time according to the accompanying clock ($\Delta t_R$), namely 
$2\Delta l_0/\Delta t_R =c$.  So the speed $v_R$ of the observer through the quantum foam  is not revealed by this
procedure, and the observer is erroneously led to the conclusion that the speed of light is always c. 
This follows from two or more observers in manifest relative motion all obtaining the same speed c by this
procedure. Despite this failure  this special effect is actually the basis of the spacetime Einstein measurement
protocol. That this protocol is blind to the absolute motion has led to enormous confusion within physics.
Later we shall see how to overcome the `blindness' of this procedure, and so manifestly reveal an observer's
$v_R$.

To be explicit the Einstein measurement protocol actually inadvertently uses this special effect by using the radar
method for assigning historical spacetime coordinates to an event: the observer records the time of emission and
reception of radar pulses ($t_r > t_e$) travelling through the space of quantum foam, and then retrospectively
assigns the time and distance of a distant event
$B$ according to (ignoring directional information for simplicity) 
\begin{equation}T_B=\frac{1}{2}(t_r+t_e), \mbox{\ \ \ }
D_B=\frac{c}{2}(t_r-t_e),\label{eqn:25}\end{equation}
  where each observer is now using the same numerical value of $c$.
 The event $B$ is then plotted as a point in 
an individual  geometrical construct by each  observer,  known as a spacetime record, with coordinates $(D_B,T_B)$. This
is no different to an historian recording events according to  some agreed protocol.  Unlike historians, who
don't confuse history books with reality, physicists do so. 
  We now show that because of this
protocol and the quantum foam dynamical effects, observers will discover on comparing their
historical records of the same events that the expression
\begin{equation}
 \tau_{AB}^2 =   T_{AB}^2- \frac{1}{c^2} D_{AB}^2,
\label{eqn:26}\end{equation}
is an invariant, where $T_{AB}=T_A-T_B$ and $D_{AB}=D_A-D_B$ are the differences in times and distances
assigned to events $A$ and
$B$ using the Einstein measurement protocol (\ref{eqn:25}), so long as both are sufficiently small
compared with the scale of inhomogeneities  in the velocity field. 

\begin{figure}[ht]
\vspace{-20mm}
\hspace{10mm}
\setlength{\unitlength}{2.0mm}
\hspace{30mm}\begin{picture}(40,45)
\thicklines
\put(-2,-2){{\bf $A$}}
\put(+1,30){{\bf $P(v_0=0)$}}
\put(16,14){\bf $B$ $(t^\prime_B)$}
\put(25,-3){\bf $D$}
\put(15,-3){\bf $D_B$}
\put(16,-0.5){\line(0,1){0.9}}
\put(-5,15){\bf $T$}
\put(26,24){$ P^\prime(v^\prime_0$)}

\put(0,0){\line(1,0){35}}
\put(0,0){\line(0,1){35}}
\put(0,35){\line(1,0){35}}
\put(35,0){\line(0,1){35}}
\put(0,0){\line(1,1){25}}

\put(0,8){\vector(2,1){16}}
\put(16,16){\vector(-2,1){16}}

\put(-3,8){\bf $t_e$}\put(-0.5,8){\line(1,0){0.9}}
\put(-3,16){\bf $T_B$}\put(-0.5,16){\line(1,0){0.9}}
\put(-3,24){\bf $t_r$}\put(-0.5,24){\line(1,0){0.9}}
\put(6,12){$\gamma$}
\put(6,22){$\gamma$}

\end{picture}
\vspace{5mm}
\caption{\small  Here $T-D$ is the spacetime construct (from  the Einstein measurement protocol) of a special observer
$P$ {\it at rest} wrt the quantum foam, so that $v_0=0$.  Observer $P^\prime$ is moving with speed
$v^\prime_0$ as determined by observer $P$, and therefore with speed $v^\prime_R=v^\prime_0$ wrt the quantum foam. Two light
pulses are shown, each travelling at speed $c$ wrt both $P$ and the quantum foam. As we see later these one-way
speeds for light, relative to the quantum foam, are equal by observation.  Event
$A$ is when the observers pass, and is also used to define zero time  for each for
convenience. }\label{fig:spacetime1}
\end{figure}
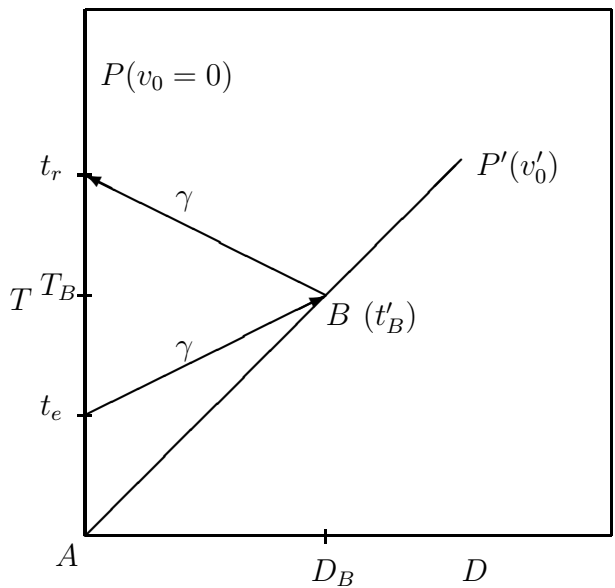

To confirm the invariant  nature of the construct in   (\ref{eqn:26}) one must pay careful attention to
observational times as distinct from protocol times and distances, and this must be done separately for each
observer.  This can be tedious.  We now  demonstrate this for the situation illustrated in
Fig.\ref{fig:spacetime1}. 

 By definition  the speed of
$P^\prime$ according to
$P$ is
$v_0^\prime =D_B/T_B$ and so
$v_R^\prime=v^\prime_0$,  where 
$T_B$ and $D_B$ are the protocol time and distance for event $B$ for observer $P$ according to
(\ref{eqn:25}).  Then using (\ref{eqn:26})  $P$ would find that
$(\tau^P_{AB})^2=T_{B}^2-\frac{1}{c^2}D_B^2$ since both
$T_A=0$ and $D_A$=0, and whence $(\tau^{P}_{AB})^2=(1-\frac{v_R^{\prime 2}}{c^2})T_B^2=(t^\prime_B)^2$ where
the last equality follows from the time dilation effect on the $P^\prime$ clock, since $t^\prime_B$ is the time
of event
$B$ according to that clock. Then $T_B$ is also the time that $P^\prime$  would compute for event $B$ when
correcting for the time-dilation effect, as the speed $v^\prime_R$ of $P^\prime$ through the quantum foam is
observable by $P^\prime$.  Then $T_B$ is the `common time' for event $B$ assigned by both
observers\footnote{Because of  gravitational in-flow effects this `common time' is not the same as a
`universal' or `absolute time'; see later. }. For
$P^\prime$ we obtain  directly, also from  (\ref{eqn:25}) and (\ref{eqn:26}), that
$(\tau^{P'}_{AB})^2=(T_B^\prime)^2-\frac{1}{c^2}(D^\prime_B)^2=(t^\prime_B)^2$, as $D^\prime_B=0$  and
$T_B^\prime=t^\prime_B$. Whence for this situation
\begin{equation}
(\tau^{P}_{AB})^2=(\tau^{P'}_{AB})^2,
\label{eqn:invariant1}
\end{equation} and so the
 construction  (\ref{eqn:26})  is an invariant.  

While so far we have only established the invariance of the construct  (\ref{eqn:26}) when one of the
observers is at rest wrt to the quantum foam, it follows that for two observers $P^\prime$ and
$P^{\prime\prime}$ both in motion wrt the quantum foam it follows that they also agree on the invariance
of (\ref{eqn:26}).  This is easily seen by using the intermediate step of  a stationary observer $P$:
\begin{equation}
(\tau^{P'}_{AB})^2=(\tau^{P}_{AB})^2=(\tau^{P''}_{AB})^2.
\label{eqn:invariant2}
\end{equation}
Hence the protocol and Lorentzian effects result in the construction in (\ref{eqn:26})  being indeed an
invariant in general.  This  is  a remarkable and subtle result.  For Einstein this invariance was a
fundamental assumption, but here it is a derived result, but one which is nevertheless deeply misleading.
Explicitly indicating  small quantities  by $\Delta$ prefixes, and on comparing records retrospectively, an
ensemble of nearby observers  agree on the invariant
\begin{equation}
\Delta \tau^2=\Delta T^2-\frac{1}{c^2}\Delta D^2,
\label{eqn:31}\end{equation} 
for any two nearby events.  This implies that their individual patches of spacetime records may be mapped one
into the other merely by a change of coordinates, and that collectively the spacetime patches  of all may
be represented by one pseudo-Riemannian manifold, where the choice of coordinates for this manifold is
arbitrary, and we finally arrive at the invariant 
\begin{equation}
\Delta\tau^2=g_{\mu\nu}(x)\Delta x^\mu \Delta x^\nu,
\label{eqn:inv}\end{equation} 
with $x^\mu=\{T,D_1,D_2,D_3\}$. 

\subsection{ The Origins of General Relativity \label{subsection:theoriginsofgeneral}} 

Above it was seen that the Lorentz symmetry of the spacetime construct  would arise if the quantum foam system that
forms space affects the rods and clocks used by observers in the manner indicated.  The effects of absolute motion
with respect to this quantum foam are in fact easily observed, and so the velocity ${\bf v}_R$ of each observer is
measurable.  However if we work only with the spacetime construct then the effects of the absolute motion are
hidden.  Einstein was very much  misled by the reporting of the experiment by Michelson and Morley of 1887, as now
(see later) it is apparent that this experiment, and others since then,  revealed evidence of absolute motion.  The
influence of the Michelson-Morley experiment had a major effect on the subsequent development of physics.  One such
development was the work of Hilbert and Einstein  in finding an apparent generalisation of Newtonian gravity to take
into account the apparent absence of absolute motion.  Despite the deep error in this work the final formulation,
known as General Relativity, has had a number of successes  including the perihelion precession of mercury, the
bending of light and gravitational red shift. Hence despite the incorrect treatment of absolute motion the formalism
of  general relativity  apparently has some validity.  In the next section we shall {\it deconstruct} this formalism
to discover its underlying physics, but here we first briefly outline the GR formalism.

 The spacetime construct  is  a static geometrical non-processing historical
record, and is nothing more than a very refined history book, with the shape of the manifold encoded in a metric
tensor $g_{\mu\nu}(x)$.   However in a formal treatment by Einstein the SR formalism and later the GR formalism is
seen to arise from three
 fundamental assumptions:
\begin{eqnarray}
&(1)& \mbox{{\bf The laws of physics have the same form in all inertial}} \nonumber\\
& &    \mbox{{\bf reference frames.}}\nonumber\\
&(2)& \mbox{{\bf Light propagates  through empty space with a definite }} \nonumber\\
& & \mbox{{\bf speed c independent of
the speed of the source or observer.}} \nonumber\\
&(3)& \mbox{{\bf In the limit of low speeds the new formalism should}} \nonumber\\
& & \mbox{{\bf agree with  Newtonian gravity. }}
\label{eqn:EinstPost}\end{eqnarray}

There is strong evidence that all three of these assumptions are in fact wrong, see \cite{RCPP2003, RCAMGE}.
Nevertheless  there is something that is partially correct within the formalism, and that part needs to be 
extracted and saved, with the rest discarded.
From the above assumptions Hilbert and Einstein
guessed the equation which specify the  metric tensor $g_{\mu\nu}(x)$, namely the geometry of the spacetime
construct,
\begin{equation}
G_{\mu\nu}\equiv R_{\mu\nu}-\frac{1}{2}Rg_{\mu\nu}=\frac{8\pi G}{c^2} T_{\mu\nu},
\label{eqn:32}\end{equation}
where  $G_{\mu\nu}$ is known as the Einstein tensor, $T_{\mu\nu}$ is the  energy-momentum tensor,
$R_{\mu\nu}=R^\alpha_{\mu\alpha\nu}$ and
$R=g^{\mu\nu}R_{\mu\nu}$ and
$g^{\mu\nu}$ is the matrix inverse of $g_{\mu\nu}$. The curvature tensor is
\begin{equation}
R^\rho_{\mu\sigma\nu}=\Gamma^\rho_{\mu\nu,\sigma}-\Gamma^\rho_{\mu\sigma,\nu}+
\Gamma^\rho_{\alpha\sigma}\Gamma^\alpha_{\mu\nu}-\Gamma^\rho_{\alpha\nu}\Gamma^\alpha_{\mu\sigma},
\label{eqn:curvature}\end{equation}
where $\Gamma^\alpha_{\mu\sigma}$ is the affine connection
\begin{equation}
\Gamma^\alpha_{\mu\sigma}=\frac{1}{2} g^{\alpha\nu}\left(\frac{\partial g_{\nu\mu}}{\partial x^\sigma}+
\frac{\partial g_{\nu\sigma}}{\partial x^\mu}-\frac{\partial g_{\mu\sigma}}{\partial x^\nu} \right).
\label{eqn:affine}\end{equation}
In this formalism the trajectories of test objects are determined by
\begin{equation}
\Gamma^\lambda_{\mu\nu}\frac{dx^\mu}{d\tau}\frac{dx^\nu}{d\tau}+\frac{d^2x^\lambda}{d\tau^2}=0,
\label{eqn:33}\end{equation}
 which is equivalent to minimising the functional
\begin{equation}
\tau[x]=\int dt\sqrt{g^{\mu\nu}\frac{dx^{\mu}}{dt}\frac{dx^{\nu}}{dt}},
\label{eqn:path}\end{equation}
wrt to the path $x[t]$.  

For the case of a spherically symmetric mass a  solution of (\ref{eqn:32}) for $g_{\mu\nu}$ outside of that mass $M$
is the Schwarzschild metric
\begin{equation}
d\tau^2=(1-\frac{2GM}{c^2r})dt^{ 2}-
\frac{1}{c^2}r^{ 2}(d\theta^2+\sin^2(\theta)d\phi^2)-\frac{dr^{ 2}}{c^2(1-\frac{\displaystyle
2GM}{\displaystyle c^2r})}.
\label{eqn:SM}\end{equation}
This solution is the basis of various experimental checks of General Relativity in which the spherically symmetric
mass is either the Sun or the Earth.  The four tests are: the gravitational redshift, the bending of light, the
precession of the perihelion of Mercury, and the time delay of radar signals.

 However the solution (\ref{eqn:SM}) is in fact
completely equivalent to the in-flow interpretation of Newtonian gravity.  Making the change of variables
$t\rightarrow t^\prime$ and
$\bf{r}\rightarrow {\bf r}^\prime= {\bf r}$ with
\begin{equation}
t^\prime=t+
\frac{2}{c}\sqrt{\frac{2GMr}{c^2}}-\frac{4GM}{c^2}\mbox{tanh}^{-1}\sqrt{\frac{2GM}{c^2r}},
\label{eqn:37}\end{equation}
the Schwarzschild solution (\ref{eqn:SM}) takes the form
\begin{equation}
d\tau^2=dt^{\prime 2}-\frac{1}{c^2}(dr^\prime+\sqrt{\frac{2GM}{r^\prime}}dt^\prime)^2-\frac{1}{c^2}r^{\prime
2}(d\theta^{\prime 2}+\sin^2(\theta^\prime)d\phi^{\prime}),
\label{eqn:PG}\end{equation}
which is exactly  the  Panlev\'{e}-Gullstrand form of the metric $g_{\mu\nu}$  \cite{PP, AG} in (\ref{eqn:24})
 with the velocity field given exactly  by the Newtonian form in (\ref{eqn:vfield}).   In which case the trajectory
equation (\ref{eqn:33}) of test objects in the Schwarzschild metric is equivalent to solving (\ref{eqn:G6}). Thus the
minimisation  (\ref{eqn:path}) is equivalent to that of  (\ref{eqn:f4}).  This choice of coordinates corresponds to
a particular frame of reference in which the test object has velocity ${\bf v}_R={\bf v}-{\bf v}_0$ relative to
the in-flow field ${\bf v}$, as seen in (\ref{eqn:f4}).  

 It is conventional wisdom for practitioners in  General
Relativity  to regard the choice of coordinates or frame of reference to be entirely arbitrary and having no physical
significance:  no observations should be possible that can detect and measure ${\bf v}_R$.  This `wisdom' is based
on two  beliefs (i) that all attempts to detect ${\bf v}_R$, namely the detection of absolute motion, have
failed, and that (ii)  the existence of absolute motion is incompatible with the many successes of both the
Special Theory of Relativity and of the General Theory of Relativity.  Both of these beliefs are demonstrably false. 

The results in this section suggest, just as for Newtonian
gravity, that the Einstein General Relativity is nothing more than the dynamical equations for a velocity flow field 
${\bf v}({\bf r },t)$.  Hence  the non-flat spacetime construct appears to be merely an unnecessary  artifact of
the Einstein measurement protocol, which in turn was motivated by the mis-reporting of the results of the
Michelson-Morley experiment. The successes of General Relativity should thus be considered as an insight into  the fluid
flow dynamics of the quantum foam system, rather than any confirmation of the validity of the spacetime formalism.  In
the next section we shall deconstruct  General Relativity to extract a possible form for this dynamics. 

\subsection{  Deconstruction of General Relativity
\label{subsection:deconstruction}}

Here we deconstruct the formalism of General Relativity by removing the obscurification produced by the
unnecessarily restricted Einstein measurement protocol.  To do this we adopt the  Panlev\'{e}-Gullstrand form  of the
metric
$g_{\mu\nu}$ as that corresponding to the observable quantum foam system, namely to an observationally detected
special frame of reference.  This form for the metric involves a general velocity field ${\bf v}({\bf r},t)$ 
where for precision we consider the coordinates ${\bf r},t$  as that of observers at rest with respect to the CMB
frame.  Note that in this frame  ${\bf v}({\bf r},t)$ is not necessarily zero, for mass acts as a sink for the flow. 
We therefore merely substitute the metric 
\begin{equation}
d\tau^2=g_{\mu\nu}dx^\mu dx^\nu=dt^2-\frac{1}{c^2}(d{\bf r}(t)-{\bf v}({\bf r}(t),t)dt)^2,
\label{eqn:PGmetric}\end{equation}
into (\ref{eqn:32})  using  (\ref{eqn:affine})  and (\ref{eqn:curvature}). This metric involves the arbitrary
time-dependent velocity field  ${\bf v}({\bf r},t)$. This is a very tedious computation and the results below were
obtained  by using the symbolic mathematics capabilities of {\it Mathematica}.  The various components of the
Einstein tensor are then
\begin{eqnarray}\label{eqn:G}
G_{00}&=&\sum_{i,j=1,2,3}v_i\mathcal{G}_{ij}
v_j-c^2\sum_{j=1,2,3}\mathcal{G}_{0j}v_j-c^2\sum_{i=1,2,3}v_i\mathcal{G}_{i0}+c^2\mathcal{G}_{00}, 
\nonumber\\ G_{i0}&=&-\sum_{j=1,2,3}\mathcal{G}_{ij}v_j+c^2\mathcal{G}_{i0},   \mbox{ \ \ \ \ } i=1,2,3.
\nonumber\\ G_{ij}&=&\mathcal{G}_{ij},   \mbox{ \ \ \ \ } i,j=1,2,3.
\end{eqnarray}
where the  $\mathcal{G}_{\mu\nu}$ are  given\footnote{The  $\mathcal{G}_{\mu\nu}$ also arise in the tetrad formulation of GR
\cite{tetrad}.} by
\begin{eqnarray}\label{eqn:GT}
\mathcal{G}_{00}&=&\frac{1}{2}((trD)^2-tr(D^2)), \nonumber\\
\mathcal{G}_{i0}&=&\mathcal{G}_{0i}=-\frac{1}{2}(\nabla\times(\nabla\times{\bf v}))_i,   \mbox{ \ \ \ \ }
i=1,2,3.\nonumber\\ 
\mathcal{G}_{ij}&=&
\frac{d}{dt}(D_{ij}-\delta_{ij}trD)+(D_{ij}-\frac{1}{2}\delta_{ij}trD)trD\nonumber\\ & &
-\frac{1}{2}\delta_{ij}tr(D^2)-(D\Omega-\Omega D)_{ij},  \mbox{ \ \ \ \ } i,j=1,2,3.
\end{eqnarray}
 Here
\begin{equation}
D_{ij}=\frac{1}{2}(\frac{\partial v_i}{\partial x_j}+\frac{\partial v_j}{\partial x_i})
\label{eqn:Dij}\end{equation}
is the symmetric  part of the rate of strain tensor $\frac{\partial v_i}{\partial x_j}$, while the antisymmetric
part is
\begin{equation}
\Omega_{ij}=\frac{1}{2}(\frac{\partial v_i}{\partial x_j}-\frac{\partial v_j}{\partial x_i}).
\label{eqn:Omegaij}\end{equation} 
In vacuum, with $T_{\mu\nu}=0$, we find from (\ref{eqn:32}) and (\ref{eqn:G}) that $G_{\mu\nu}=0$ implies that  
$\mathcal{G}_{\mu\nu}=0$. It is then easy to check that the in-flow velocity field (\ref{eqn:vfield})  satisfies
these equations.  This simply expresses the previous observation that this `Newtonian in-flow' is completely
equivalent to the Schwarzschild metric.  We note that the vacuum equations  $\mathcal{G}_{\mu\nu}=0$ do not involve
the speed of light; it appears only in (\ref{eqn:G}). It is therefore suggested that (\ref{eqn:G}) amounts to the
separation of the Einstein measurement protocol, which involves $c$, from the supposed dynamics of gravity within the  GR
formalism, and which does not involve $c$. However the details of the vacuum dynamics in (\ref{eqn:GT}) have not
actually been tested.  All the key tests of GR are now seen to amount to a test {\it only} of $\delta \tau[x]/\delta
x^\mu = 0$, which is the minimisation of  (\ref{eqn:f4}),  when the in-flow field is given by  (\ref{eqn:G}), and which
is nothing more than Newtonian gravity. Of course Newtonian gravity was itself merely based upon observations within
the Solar system, and this may have been too special to have revealed key aspects of gravity. Hence, despite popular
opinion, the GR formalism is apparently based upon  rather poor evidence.

\subsection{  Gravity as Inhomogeneous Quantum-Foam Flow \label{subsection:gravity}}
 
Despite the limited insight into gravity which GR is now seen to amount to, here we look for possible
generalisations of Newtonian gravity and its in-flow interpretation by examining some of the mathematical
structures that have arisen in  (\ref{eqn:GT}). For the case of zero vorticity
$\nabla\times{\bf v}=0$ we have
$\Omega_{ij}=0$  and also that we may write 
${\bf v}=\nabla u$ where $u({\bf r},t)$ is a scalar field, and  only one equation is required to determine $u$.
 To that end  we  consider the trace of  
$\mathcal{G}_{ij}$. Note that
$tr(D)=\nabla.{\bf v}$, 
and that
\begin{equation}
\frac{d (\nabla.{\bf v})}{dt}=({\bf v}.\nabla)(\nabla.{\bf v})+\frac{\partial (\nabla.{\bf v})}{\partial t}.
\end{equation}
Then using the identity 
\begin{equation}
({\bf v}.\nabla)(\nabla.{\bf v})=\frac{1}{2}\nabla^2({\bf v}^2)
-tr(D^2)-\frac{1}{2}(\nabla\times{\bf v})^2+ {\bf v}.\nabla\times(\nabla\times{\bf v}),
\label{eqn:identity}\end{equation}
and imposing 
\begin{equation}
\sum_{i=1,2,3}\mathcal{G}_{ii}=-8\pi G\rho,
\label{eqn:new}\end{equation}
 we obtain
\begin{equation}
\frac{\partial}{\partial t}(\nabla.{\bf v})+\frac{1}{2}\nabla^2({\bf v}^2)+\frac{1}{4}((tr D)^2-tr(D^2))=-4\pi
G\rho.
\label{eqn:newgravity}\end{equation}
This is seen to be a possible  generalisation of the Newtonian equation (\ref{eqn:f3}) that includes a
time derivative, and also the new term $C({\bf v})=\frac{1}{4}((tr D)^2-tr(D^2))$.    First note that for
the case of the Solar system, with the mass concentrated in one object, namely the Sun, we see that the
in-flow field (\ref{eqn:vfield}) satisfies (\ref{eqn:newgravity}) since then $C({\bf v})=0$. As we shall see later
the presence of the $C$ term is also well hidden when we consider the Earth's gravitational effects,
although there are various known anomalies that indicate that a generalisation of Newtonian gravity is
required. 
Hence (\ref{eqn:newgravity}) in the case of the Solar system is indistinguishable from Newtonian
gravity, or the Schwarzschild metric within the General Relativity formalism   so long as we use
(\ref{eqn:f4}), in being able to determine trajectories of test objects.  Hence  (\ref{eqn:newgravity}) is
automatically  in agreement with most of the so-called checks on Newtonian gravity and later General Relativity. 
Note that (\ref{eqn:newgravity}) does not involve the speed of light $c$. Nevertheless  we have not derived 
(\ref{eqn:newgravity})) from the underlying QHFT, and indeed it is not a consequence of GR, as 
the  $\mathcal{G}_{00}$ equation of  (\ref{eqn:GT}) requires that  $C({\bf v})=0$ in vacuum.
Eqn.(\ref{eqn:newgravity}) at this stage should be regarded as a conjecture  which will  permit the exploration of
possible quantum-flow physics and also allow comparison with experiment. 

However one key aspect of  (\ref{eqn:newgravity}) should be noted
here,  namely that being a non-linear fluid-flow dynamical system we would expect the flow to be
turbulent, particularly when the matter is not spherically symmetric  or inside even a spherically
symmetric distribution of matter, since then the $C({\bf v})$ term is non-zero and it will drive that
turbulence.  In the following sections we shall see that the experiments that reveal absolute motion also reveal
evidence of turbulence.    

 Because of the  $C({\bf v})$ term  (\ref{eqn:newgravity}) 
would predict that the Newtonian inverse square law would not be applicable  to systems
such as spiral galaxies, because of their highly non-spherical distribution of matter.  Of course
attempts to retain this law, despite its   manifest
 failure,  has   led to the  spurious introduction of  the notion of dark
matter within spiral galaxies, and also at larger scales.  
From 
\begin{equation}\label{eqn:ga}
{\bf g}=\frac{1}{2}\nabla({\bf v}^2)+\frac{\partial {\bf v}}{\partial
t},\end{equation} 
which is (\ref{eqn:f2}) for irrotational flow, we see that (\ref{eqn:newgravity})  gives
\begin{equation}\label{eqn:g2}
\nabla.{\bf g}=-4\pi G\rho-C({\bf v}),
\end{equation}
and taking running time averages to account for turbulence
\begin{equation}\label{eqn:g3}
\nabla.\!\!<\!\!{\bf g}\!\!>=-4\pi G\rho-<\!\!C({\bf v})\!\!>,
\end{equation}
and writing  the extra term as $<\!\!C({\bf v})\!\!>=4\pi G \rho_{DM}$ we see that  $\rho_{DM}$ would act as an
effective matter density, and it is suggested that it is the consequences of this term which have been
misinterpreted as `dark matter'.  Here we see that this effect is actually the consequence of quantum foam
effects within the new proposed dynamics for gravity, and which becomes apparent particularly in spiral
galaxies.  Note that (\ref{eqn:newgravity}) is an equation for
${\bf v}$, and now involves the direction of ${\bf v}$, unlike the special case of Newtonian gravity (\ref{eqn:f3}). 
Because $\nabla\times{\bf v}=0$ we can write (\ref{eqn:newgravity}) in the form
\begin{equation}\label{eqn:newint}
{\bf v}({\bf r},t)=\frac{1}{4\pi}\int^t dt^\prime\int d^3r^\prime ({\bf r}-{\bf r}^\prime)
\frac{
\frac{1}{2}\nabla^2({\bf v}^2({\bf r^\prime},t^\prime))
+4\pi G\rho({\bf r^\prime},t^\prime)
+C({\bf v}({\bf r^\prime},t^\prime))}
{|{\bf r}-{\bf r}^\prime|^3},
\end{equation}
which allows the  determination of the time evolution of ${\bf v}$.

The new flow dynamics encompassed in  (\ref{eqn:newgravity})  thus
accounts for most of the known gravitational phenomena, but will lead to some very clear cut experiments
that will distinguish it from the two previous attempts to model gravitation.  It turns out that these two
attempts were based on some key `accidents' of history. In the case of the Newtonian modelling of gravity
the prime `accident' was of course  the Solar system with its high degree of spherical symmetry.  In
each case we had test objects, namely the planets, in orbit about the Sun, or we had test object in orbit
about the Earth.  In the case of the General Relativity modelling the prime `accident' was the
mis-reporting of the Michelson-Morley experiment, and  the ongoing belief that the so called
`relativistic effects' are incompatible with absolute motion.  We shall consider in detail later some
further anomalies that might be appropriately explained by this new modelling of gravity.  Of course that
the in-flow has been present in various experimental data is also a significant argument for something like 
(\ref{eqn:newgravity}) to model gravity.  Key new experimental techniques will be introduced later which wil enable
the consequences of (\ref{eqn:newgravity}) to be tested. If necessary these  experiments will provide insights
into possible modifications to  (\ref{eqn:newgravity}).

\subsection{ In-Flow Superposition Approximation \label{subsection:inflowsuperposition}}

We consider here why the presence of the $C({\bf v})$ term appears to have escaped attention in the case of
gravitational experiments and observations near the Earth, despite the fact that the presence of the Earth breaks
the spherical symmetry of the matter distribution of   the Sun. 

First note that if we have a matter 
distribution  at rest in the space of quantum foam, and that (\ref{eqn:newgravity}) has solution ${\bf v}_0({\bf
r},t)$, then when the same matter distribution is uniformly translating at velocity ${\bf V}$, that is 
$\rho({\bf r})\rightarrow  \rho({\bf r}-{\bf V}t)$, then a solution to (\ref{eqn:newgravity}) is 
 \begin{equation}\label{eqn:Vsum}
{\bf v}({\bf r},t)={\bf v}_0({\bf r}-{\bf V}t)+{\bf V}.
\end{equation}
This follows because (i) the  expression for the acceleration ${\bf g}({\bf r},t)$ gives
\begin{eqnarray}\label{eqn:Vsum1}
{\bf g}({\bf r},t) &=&
\frac{\partial {\bf v}_0({\bf r-{\bf V}t})}{\partial t}+(({\bf v}_0({\bf
r}-{\bf V}t)+{\bf V}).{\bf \nabla)}({\bf v}_0({\bf r}-{\bf V}t)+{\bf
V}),\nonumber\\
&=&
\frac{\partial {\bf v}_0({\bf r-{\bf V}t})}{\partial t}+{\bf g}_0({\bf r-{\bf
V}t})+({\bf V}.{\bf \nabla)}{\bf v}_0({\bf r}-{\bf V}t),\nonumber\\
&=&
-({\bf
V}.{\bf
\nabla)}{\bf v}_0({\bf r}-{\bf V}t)+{\bf g}_0({\bf r-{\bf V}t})+({\bf
V}.{\bf \nabla)}{\bf v}_0({\bf r}-{\bf V}t),\nonumber\\
&=&{\bf g}_0({\bf r-{\bf V}t}),
\end{eqnarray} 
as there is a cancellation of two terms in (\ref{eqn:Vsum1}), and where  ${\bf g}_0({\bf r})$ is the acceleration
field when the matter distribution is not in translation, and (ii) clearly $C({\bf v}_0({\bf r}-{\bf V}t)+{\bf
V}) = C({\bf v}_0({\bf r}-{\bf V}t))$, and  so this term is  also simply translated.
 Hence apart from the translation effect the acceleration is the same.  Hence the velocity vector
addition rule in (\ref{eqn:Vsum}) is valid.  

Now for  Earth based  gravitational phenomena the motion of the Earth takes place within the velocity in-flow 
towards the Sun, and the velocity sum rule (\ref{eqn:Vsum})  is only approximately valid  as now ${\bf V}\rightarrow
{\bf V}({\bf r},t)$ and no longer corresponds to uniform translation, and may manifest turbulence.  To be a valid
approximation the inhomogeneity of ${\bf V}({\bf r},t)$ must be much smaller than that of  ${\bf v}_0({\bf r}-{\bf
V}t)$, which it is\footnote{The Earth's centripetal acceleration about the Sun is approximately 1/1000 that of the 
gravitational acceleration at the surface of the Earth. }.  Nevertheless  turbulence   associated  with the
$C({\bf v})$ term is apparent in experimental data.  The validity of this approximation plays a special role in
explaining why the entrainment hypotheses regarding the detection of Earth's absolute motion was unnecessary.

\subsection{  Measurements of $G$ \label{subsection:measurementsofG}}

As noted in Sect.\ref{subsection:classicalquantum} Newton's Inverse Square Law of Gravitation
may only be strictly valid in cases of spherical symmetry.  The  theory that gravitational effects
arise from inhomogeneities in the quantum foam flow  implies that there is no `universal law of
gravitation' because the inhomogeneities are determined by non-linear `fluid equations' and the
solutions  have no form which could be described by a `universal law'.  Fundamentally there is no
generic fluid flow behaviour. The Inverse Square Law is then only an approximation, with  large
deviations expected in the case of spiral galaxies. Nevertheless Newton's gravitational constant $G$
will have a definite value as it quantifies the effective rate at which matter dissipates the
information content of space.  

From these considerations it
follows that the measurement of the value of $G$ will be difficult as the measurement of the  forces
between two of more objects, which is the usual method of measuring $G$, will depend on the geometry
of the spatial positioning of these objects  in a way not previously accounted for because the
Newtonian Inverse Square Law has always been assumed, or in some case  a specified change in the
form of the law has been used.  But in all cases a `law' has been assumed, and this may have been
the flaw in  the analysis of data from such experiments.  This implies that the value of
$G$ from such experiments will show some variability as a systematic effect has  been neglected in
analysing the experimental data, for in none of these experiments is spherical symmetry present.  So
experimental measurements of $G$  should show an unexpected contextuality.  As well the influence of
surrounding matter has also not been properly accounted for. Of course  any effects of turbulence in
the inhomogeneities of the flow has presumably never even been contemplated.  

The first measurement of $G$ was in 1798 by Cavendish using a torsional balance.  As the precision of
experiments increased over the years and a variety of techniques used  the disparity between the
values of $G$ has actually increased.  In 1998 CODATA increased the uncertainty in $G$ from 0.013\% to 0.15\%. 
One indication of the contextuality is that measurements of $G$  produce values that differ by nearly
40 times their individual error estimates \cite{G1}.  It is predicted that these $G$ anomalies will only be resolved
when  the new theory of gravity is used in analysing the data from these experiments. 

\vspace{3mm}

\subsection{  Gravitational Anomalies\label{subsection:gravitationalanomalies}}

In Sect.\ref{subsection:measurementsofG} anomalies associated with the measurement of $G$ were briefly discussed and
it was pointed out that these were probably explainable within the new in-flow theory of gravity.  There are in-fact
additional gravitational anomalies that are not well-known in physics, presumably because their existence is
incompatible with the Newtonian or the  Hilbert-Einstein gravity theories.

The most significant of these anomalies is the Allais effect \cite{Allais}.   In June 1954 Allais\footnote{Maurice
Allais won the Noble Prize for Economics in 1988.} reported that a Foucault
pendulum exhibited peculiar movements at the time of a solar eclipse.  Allais was recording the precession of a
Foucault pendulum in Paris. Coincidently during the 30 day observation period a partial solar eclipse occured at
Paris on June 30.  During the eclipse the precession of the pendulum was seen to be disturbed.  Similar results were
obtained during another solar eclipse on October 29 1959.  There have been other repeats of the Allais experiment
with varying results.  

Another anomaly was reported by Saxl and Allen \cite{Saxl} during the solar eclipse of March 7 1970.  Significant
variations in the period of a torsional pendulum were observed  both during the eclipse and as well in the hours just
preceding and just following the eclipse.  The effects seem  to suggest that an ``apparent wavelike structure has been observed over the
course of many years at our Harvard laboratory'', where the wavelike structure is present and reproducible even in the
absence of an eclipse. 

Again Zhou and Huang \cite{Zhou}  report various time anomalies occuring during the solar eclipses of Sepetmber 23
1987,  March 18 1988 and  July 22 1990 observed using atomic clocks.

All these anomalies and others not discussed here would suggest that  gravity has aspects to it that are not within
the prevailing theories, but that the in-flow theory discussed above might well provide an explanation, and indeed
these anomalies may well provide further phenomena that could be used to test the new theory.  The effects  associated
with the solar eclipses could presumably follow from the alignment of the Sun, Moon and the Earth causing
enhanced turbulence.  The Saxl and Allen experiment of course suggests, like the other experiments analysed
above, that the turbulence is always present. To explore these anomalies detailed numerical studies of
(\ref{eqn:newgravity}) are required with particular emphasis on the effect on the position of the Moon.

\subsection{    Galactic In-flow  and the CMB Frame\label{subsection:galacticinflow}}  
Absolute motion (AM) of the Solar system has been observed in the direction
$(\alpha=17.5^h,\delta=65^0)$, up to an overall sign to be sorted out,  with a speed of
$417 \pm 40$ km/s \cite{RCPP2003}. This is the velocity after  removing the contribution of the Earth's
orbital speed and the Sun in-flow effect. It is significant that this velocity is different
to that associated with the Cosmic Microwave Background \footnote{The understanding of the galactic 
in-flow effect was not immediate: In
\cite{CK} the direction was not determined, though the speed was found to be comparable to 
the CMB determined speed.  In \cite{RC03} that the directions were very different was 
noted but not appreciated, and in fact thought to be due to experimental error. In
\cite{RC05} an analysis of some of the `smoother' Michelson-Morley data resulted in an 
incorrect direction. At that stage it was not understood that the data showed large 
fluctuations in the azimuth apparently caused by the turbulence. Here the issue is hopefully 
finally resolved.} 
(CMB) relative to which the Solar system
has a speed of $369$ km/s in the direction
 $(\alpha=11.20^h,\delta=-7.22^0)$, see \cite{CMB}. 
This CMB velocity is obtained by finding the preferred frame in which this thermalised
$3^0$K radiation is isotropic, that is by removing the dipole component.  
The CMB velocity is a measure of the motion
of the Solar system relative to the universe as a whole, or aleast a shell of the universe
some 15Gyrs away, and indeed the near uniformity of that radiation in all directions
demonstrates that we may  meaningfully refer to the spatial structure of the
universe.  The concept here is that at the time of decoupling of this radiation from
matter that matter was on the whole, apart from small observable fluctuations, at
rest with respect to the quantum-foam system that is space. So the CMB velocity is the
motion of the Solar system  with respect to space {\it universally},  but not
necessarily with respect to the   {\it local} space.  Contributions to this  velocity
would arise from the orbital motion of the Solar system within the Milky Way galaxy,
which has  a speed of some 250 km/s, and contributions from the motion of the Milky
Way within the local cluster, and so on to perhaps larger clusters.

On the other hand the AM velocity is a vector sum of this {\it universal} CMB
velocity and the net velocity associated with the {\it local} gravitational in-flows into
the Milky Way and the local cluster.  If the CMB velocity had been identical to the AM
velocity then the in-flow  interpretation of gravity would have been proven wrong. We
therefore have three pieces of experimental evidence for this interpretation (i) the
refractive index anomaly discussed previously in connection with the Miller data, (ii) the
turbulence seen in all detections of absolute motion, and now (iii) that the AM velocity is
different in both magnitude and direction from that of the  CMB  velocity, and that this
velocity does not display the turbulence seen in the AM velocity. 

That the AM and CMB velocities are different amounts to the discovery of the resolution 
to the `dark matter' conjecture. Rather than the galactic velocity anomalies being caused by
such undiscovered `dark matter' we see that the in-flow into non spherical galaxies, such as
the spiral Milky Way, will be non-Newtonian \cite{RCPP2003}.   As well it will be interesting to
determine, at least theoretically, the scale of turbulence expected in galactic systems,
particularly as the magnitude of the turbulence seen in the AM velocity is somewhat
larger than might be expected from the Sun in-flow alone. Any theory for the turbulence
effect will certainly be checkable within the Solar system as the time scale of this
is suitable for detailed observation.

It is also clear that the time of obervers  at rest with respect to the CMB frame is 
absolute or  universal time.  This interpretation of 
the CMB frame has of course always been rejected by supporters of the SR/GR formalism.
As for space we note that it has a differential structure, in that different regions
are in relative motion.  This is caused by the gravitational in-flow effect locally, and 
as well by the growth of the universe.

\subsection{  In-Flow Turbulence and Gravitational Waves \label{subsection:inflowturbulence}}

The velocity  flow-field equation (\ref{eqn:newgravity}) is
expected to have solutions possessing turbulence, that is, random fluctuations in both the
magnitude and direction of the gravitational in-flow component of the velocity flow-field.   
Indeed all the Michelson interferometer experiments showed evidence of such turbulence. The first
clear evidence was from the Miller experiment, as shown discussed in \cite{RCPP2003}.  Miller offered no
explanation for these fluctuations  but in his analysis of that data he did running time averages.  
Miller may have in fact have simply interpreted these fluctuations as purely instrumental effects.  While
some of these fluctuations may be partially caused by weather related temperature and pressure 
variations, the bulk of the fluctuations appear to be larger than expected from that cause alone.   Even
the original Michelson-Morley data, plotted in \cite{RCPP2003} shows variations in  the velocity field and
supports this interpretation.    However it is significant that the non-interferometer DeWitte
\cite{RCPP2003} data also shows evidence of turbulence in both the magnitude and  direction of the velocity
flow field.  Just as the DeWitte data agrees
with the Miller data for speeds and directions the magnitude fluctuations are very similar in absolute
magnitude as well.  

It therefore  becomes clear that there is
strong evidence for these fluctuations being evidence of  physical turbulence in the flow
field.  The magnitude of this turbulence appears to be somewhat larger than that which would be
caused by the in-flow of quantum foam towards the Sun, and indeed following on from
Sect.\ref{subsection:galacticinflow}  some of this turbulence may be associated with galactic
in-flow into the Milky Way.  This in-flow turbulence is a form of gravitational wave and the
ability of gas-mode Michelson interferometers to detect absolute motion means that experimental
evidence of such a wave phenomena has been available for a considerable period of time, but
suppressed along with the detection of absolute motion itself.   Of course flow equations  do not exhibit those
gravitational waves of the form that have been predicted to exist based on the Einstein equations, and which are
supposed to propagate at the speed of light.  All this means that gravitational wave phenomena is very easy to detect
and amounts to new physics that can be studied in much detail, particularly using the new 1st-order
interferometer discussed in \cite{RCPP2003}.

\subsection{  Absolute Motion and Quantum Gravity\label{subsection:absolutmeqg}}
 
Absolute rotational motion had been recognised as a meaningful and obervable phenomena from the very beginning of
physics. Newton had used his rotating bucket experiment to illustrate the reality of absolute rotational motion,
and later Foucault  and Sagnac provided further experimental proof. But for absolute linear motion the history
would turm out to be completely different. It was generally thought that absolute linear motion was undetectable,
at least until Maxwell's electromagnetic theory appeared to require it. In perhaps the most bizarre sequence of
events in modern science it turns out that absolute linear motion has been apparent within experimental data for
over 100 years. It was missed in the first experiment designed to detect it and from then on for a variety of
sociological reasons it became a concept rejected  by physicists and banned from their journals despite
continuing  new experimental evidence. Those who pursued the scientific evidence were treated with scorn and
ridicule.  Even worse was  the impasse that this obstruction of the scientific process resulted in, namely the
halting of nearly all progress in furthering our understanding of the phenomena of gravity. For it is clear from
all the experiments that were capable of detecting absolute motion that there is present in that data evidence of
turbulence within the velocity field.  Both the in-flow itself and the  turbulence are manifestations at a
classical level of what is essentially  quantum gravity processes. 

Process Physics has given a unification of explanation and description of physical phenomena based upon the limitations of
formal syntactical systems which had nevertheless achieved a remarkable encapsulation of many phenomena, albeit in a
disjointed and confused manner, and with a dysfunctional ontology  attached for good measure.  As argued in
\cite{RCPP2003, RCAMGE} space is a quantum system continually classicalised by on-going non-local collapse
processes.  The emergent phenomena is foundational to existence and experientialism.   Gravity in this system is
caused by  differences in the
 rate of processing of the cellular information within the network which we experience as space, and
consequentially there is a differential flow of information which can be affected by the presence of matter or even by
space itself.  Of course the motion of matter including photons with respect to that spatial information content  is
detectable because it affects the geometrical and chronological attributes of that matter, and the experimental evidence
for this has been exhaustively discussed in \cite{RCPP2003, RCAMGE}.   What has become very clear is that the
phenomena of gravity is only understandable once we have this unification of the quantum phenomena of matter and
the quantum phenomena of space itself. In Process Physics the difference between matter and space is subtle. It
comes down to the difference between informational patterns that are topologically preserved and those
information patterns that are not.    One outcome of this unification is that as a consequence of having a
quantum phenomena of space itself we obtain an informational explanation for gravity, and which at a suitable
level  has an emergent quantum description.  In this sense we have an emergent quantum theory of gravity.  Of
course no such quantum description of gravity  is derivable from quantising Einsteinian gravity itself. This
follows on two counts, one is that the Einstein gravity formalism  fails on several levels, as discussed
previously, and second that quantisation has no validity as a means of uncovering deeper physics.   Most
surprising of all is that   having uncovered the logical necessity for gravitational phenomena it also appears
that even the seemingly well-founded Newtonian account of gravity has major failings.  The denial of this
possibility has resulted in an unproductive search for dark matter.  Indeed like dark matter and spacetime much
of present day physics has all the hallmarks of another episode of Ptolemy's epicycles, namely concepts that
appear to be well founded but in the end turn out to be illusions, and ones that have acquired the status of
dogma.

If the Michelson-Morley experiment had been properly analysed and the phenomena revealed by the data exposed, and this
would have required in 1887 that Newtonian physics be altered, then as well as the subsequent path of physics being very
different, physicists would almost certainly have discovered both the gravitational in-flow effect and associated  
 turbulence.  

It is clear then that observation and measurement of absolute motion leads directly to a
changed paradigm regarding the nature and manifestations of gravitational phenomena, and that the new 1st-order
interferometer described in \cite{RCPP2003}  will provide an extremely simple device to uncover
aspects of gravity previously denied by current physics.  There are two aspects of such an experimental program. One is
the characterisation of the turbulence and its linking to the new non-linear term in the velocity field theory. This is
a top down program. The second aspect is a bottom-up approach where the form of the velocity field theory, or its
modification, is derived from the deeper informational process physics.  This is essentially the quantum gravity
route.   The turbulence is of course essentially a gravitational wave phenomena and networks of 1st-order
interferometers will permit  spatial and time series analysis.  There are a number of other  gravitational anomalies
which may also  now be studied using such an interferometer network, and so much new physics can be expected to be
uncovered.

\section{ Conclusions\label{section:conclusions}}

 Here  a new theory of gravity has been proposed.  It passes all the key existing tests, and also appears
to be capable of explaining numerous gravitational anomalies.   The phenomena present in these anomalies
provide opportunities for further tests of the new gravitational physics.  This new theory is supported
by the Miller abolute motion experiment in that it reveals the turbulent in-flow of space associated with
gravity, as well as the existence of absolute motion itself. This clearly refutes the fundamental
postulates of the Einstein reinterpretation of the relativitsic effects that had been developed by Lorentz
and others.  Indeed these experiments are consistent with the Lorentzian relativity in which reality
displays both absolute motion effects {\it and} relativistic effects. As discussed in detail in
\cite{RCPP2003} it is absolute motion that actually causes these relativistic effects.  Both General
Relativity and the Newtonian theory, for which GR was constructed to agree with in the low speed limit, are
refuted by these experiments. These developments are discussed more extensively in \cite{RCPP2003}.  As
well the dynamical structure of stars needs to be reexamined in the light of this new theory, and this may
have some relevance to the neutrino flux problem.

\section{ References\label{section:references}}

\end{document}